\documentclass{aa}
\usepackage{natbib,psfig,graphicx}
\def\aproxsim{\mbox{$\begin{array}{c}
  \propto\\[-1.6ex]
  \sim \end{array}$}}
\def\nodata{ ~$\cdots$~ }

\begin{document}


\title{An XMM-Newton view of the cluster of galaxies Abell~85}

\author{F. Durret\inst{1} \and G.B. Lima Neto\inst{2} \and 
W. Forman\inst{3}}

\institute{
Institut d'Astrophysique de Paris, CNRS, 98bis Bd Arago, 75014 Paris, France 
\and
Instituto de Astronomia, Geof\'{\i}sica e C. Atmosf./USP, R. do Mat\~ao 1226, 
05508-090 S\~ao Paulo/SP, Brazil 
\and
Harvard Smithsonian Center for Astrophysics, 60 Garden St, Cambridge MA
02138, USA}

\date{Accepted ??/2004. Received ??/2004}

\authorrunning{Durret et al.}
\titlerunning{XMM-Newton view of Abell~85}

\abstract{We have observed the cluster of galaxies Abell 85 with
XMM-Newton.  These data have allowed us to confirm in a previous paper
the existence of the extended 4~Mpc filament detected by the ROSAT
PSPC in the neighbourhood of this cluster, and to determine an X-ray
temperature of about $\sim 2\,$keV. We now present a thorough analysis
of the properties of the X-ray gas in the cluster itself, including
temperature and metallicity maps for the entire cluster. These results
show that Abell~85 had intense merging activity in the past and is not
fully relaxed, even in the central region. We have also determined the
individual abundances for some iron-group metals and $\alpha$-elements
in various regions; the ratios of these metallicities to the iron
abundance show that both supernova types Ia and II must be involved
in the intra-cluster gas enrichment. Spectral analysis of the central
region suggests a different redshift of the X-ray emitting gas
compared to the mean cluster velocity derived from galaxy member
redshifts. We discuss the implications of the difference between the
cD galaxy redshift, the mean galaxy redshift and the hot gas redshift,
as well as the possibility of several groups being accreted on to
Abell~85. Finally, we obtain the dynamical mass profile and baryon
fraction taking into account the new determined temperature profile.
The dynamical mass in Abell~85 has a steep density profile, similar to
the ones found in $N$-body simulations.  }

\maketitle

\section{Introduction}\label{sec:intro}

Clusters of galaxies are the largest and latest bound, relaxed structures to
form in the Universe. They are used to probe large scale structure and
evolution of galaxies in dense environments. A deep understanding of clusters
is therefore necessary if they are to be effectively used as an astrophysical
tool and laboratory, and detailed multi-wavelength observations of nearby
clusters are one of the best means to reach a better knowledge of clusters.

The complex of clusters Abell~85/87/89 is a very well studied structure at
both X-ray [with ROSAT, \citet{Pislar97,LimaNeto97}, and with BeppoSAX,
\citet{LimaNeto01}], and optical wavelengths, with extensive redshift and
imaging catalogues \citep{Durret98a,Slezak98}; the mean galaxy redshift of
Abell~85 is $z=0.0555$ \citep{Durret98b}.\footnote{At $z=0.0555$, 1 arcmin $=
90.5 h_{50}^{-1}\,$kpc, assuming $\Omega_M=0.3$, $\Omega_\Lambda=0.7$.} As
shown by \cite{Durret98b}, galaxies in the region of Abell~87 have redshifts
comparable to those in Abell~85, but Abell~87 is not detected in X-rays (as
confirmed by the present XMM-Newton data), suggesting that it is not a
separate cluster or group. Abell~89 comprises two background structures and is
not clearly detected in X-rays either, so we will not consider these two
``Abell clusters'' any further.

The general picture drawn from a combined X-ray (ROSAT PSPC) and optical
(imaging and large redshift catalogue) analysis \citep{Durret98b} and
references therein, shows the existence of a main cluster, of a south blob
which is probably a group at the same redshift as the main cluster, and of an
extended filament (at least 4~Mpc long). The latter may be diffuse emission or
a chain of several groups of galaxies, which seem to be falling on to the main
cluster, the impact region being located in projection somewhat north of the
south blob.

In order to analyse better the morphology and physical properties of the X-ray
gas in various regions of the main cluster and in the filament, we have
observed Abell~85 with XMM-Newton. Our first results concerning the filament
were described by \citet{LimaNeto02,Durret03}. We present here the results
concerning the physical properties of the X-ray gas in the main cluster. They
are complementary to those of \citet{Kempner02}, who discuss the Chandra
observation of Abell 85 with emphasis on the X-ray concentration at the
northernmost end of the filamentary structure: the south subcluster or south
blob (the filament lies outside their field).

\begin{figure*}[!htb]
\centering 
\includegraphics[width=12cm]{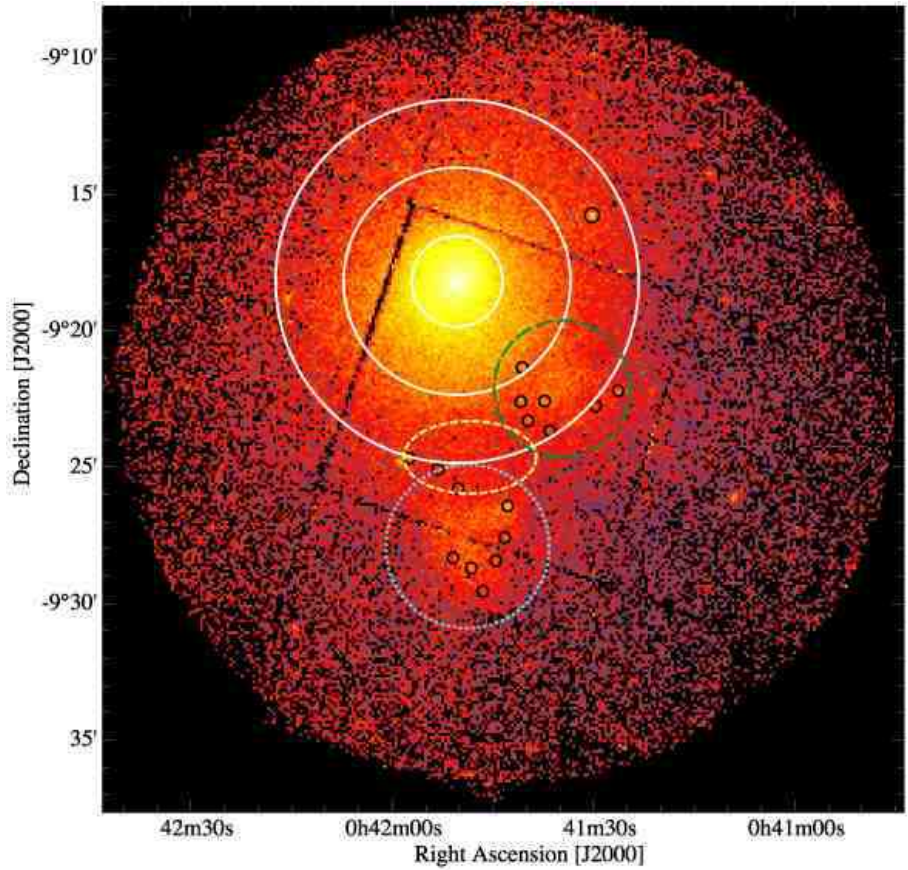}
\caption[]{XMM MOS1 and MOS2 raw image of Abell 85, with the regions analysed
in detail superimposed. The concentric circles around the cluster centre
(\textsl{white full lines}) define the centre and rings used for the detailed
spectral analysis; the \textsl{green dashed} circle defines the VSSRS
region; the \textsl{yellow dashed} ellipse defines the impact region,
where the South Blob is probably hitting the main body of Abell~85; the
\textsl{blue dotted} circle defines the South Blob region. The \textsl{small
black circles} are point sources that were excluded when accumulating the
spectra. The VSSRS region was also excluded from the external ring.}
\label{fig:Xray-all}
\end{figure*}

\begin{figure*}[!htb]
\centering
\includegraphics[width=13cm]{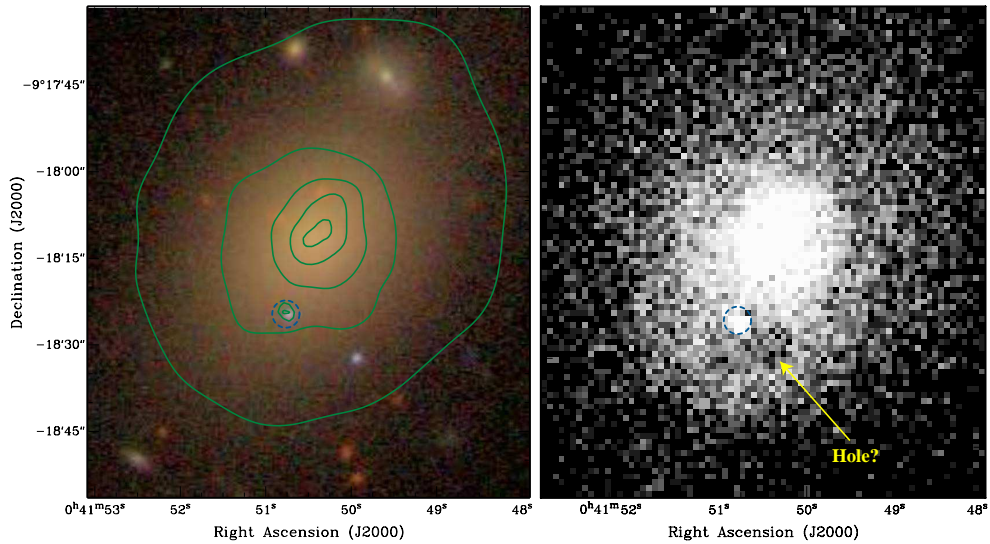} 
\caption[]{\textsl{Left:} Optical``true-color'' \textit{r-g-i} image
from the SDSS 2nd Data Release overlayed with logarithmically spaced
Chandra isocontours. \textsl{Right:} Chandra X-ray image in the
[0.3--8.0 keV] band. The small circle on both panels indicates an
X-ray point source which may be an AGN. The arrow at the bottom right
shows the position of a ``hole'' of X-ray emission south of the
nucleus clearly visible on the Chandra image.}
\label{fig:optX}
\end{figure*}

\section{Observations and data reduction}\label{sec:obs}

Two XMM-Newton \citep{Jansen01} observations were performed on January
7th, 2002. The first exposure was pointed between the center of
Abell~85 (coincident with the cD galaxy) and the south blob; the
second observation was aimed at the southern filament.  Both total
exposure times were 12.5~ks using the \textsc{medium} optical filter
in standard Full Frame mode. The basic data processing (the
``pipeline'' removal of bad pixels, electronic noise and correction
for charge transfer losses) was done with SAS V5.3. After the
paper was submitted, SAS V6.0 became available and we tested that
several of our results (the temperature profile, the South blob
spectral fits) remained within the error bars, thus making a new
analysis unnecessary.

For the spectral analysis, we have used the observations made with the
EPIC/MOS1, MOS2 and PN cameras. For the EPIC/MOS cameras, after applying the
standard filtering, keeping only events with FLAG~$=0$ and PATTERN $\le$ 12,
we have removed the observation times with flares using the light-curve in the
[10--12 keV] band using a 3-sigma clip technique. The cleaned MOS1 and MOS2
event files have remaining exposure times of $12\;381$~s and $12\;386$~s,
respectively.

We did similarly for the PN, applying the standard filtering and keeping only
events with FLAG~$=0$ and PATTERN $\le$ 4. The cleaned PN event file has a
useful exposure time of $9\;039$~s.

The redistribution and ancillary files (RMF and ARF) were created with
the SAS tasks \texttt{rmfgen} and \texttt{arfgen} for each camera and each
region that we analysed.

The smoothed, merged image of the four cleaned MOS observations in the
0.3--5.0~keV energy band has already been shown by \citet{Durret03},
confirming our previous detection of the ``filament'' with the ROSAT
PSPC.

Spectra were analyzed with \textsc{xspec}~11.3. Since the spectra were
rebinned, we have used standard $\chi^{2}$ minimization. We have
applied both \textsc{mekal} \citep{Kaastra93,Liedahl95} and
\textsc{apec} (Chandra X-Ray Center, \textsc{atomdb} Version 1.3.1)
plasma models.

The metal abundances (or metallicities) are based on the solar values given by
\citet{Anders89}. We chose to use these abundances for easier comparison with
previous work, even though these values may not be accurate: more recent solar
abundance determinations \cite[e.g.][]{Grevesse98,Wilms00} give
significatively lower O and Fe abundances than \citet{Anders89}. This is
important to take into account when we use ``variable metallicity'' models in
order to fit the individual metal abundances (the \textsc{vmekal} and
\textsc{vapec} plasma models).

The photoelectric absorption -- mainly due to neutral hydrogen in our galaxy
-- was computed using the cross-sections given by \cite{Balucinska92},
available in \textsc{xspec}.

\section{Global properties of the X-ray gas}

\subsection{Overall morphology}

The overall XMM-Newton image of Abell~85 is displayed in
Fig.~\ref{fig:Xray-all}, with the various regions analyzed in detail
below indicated. The main cluster, South Blob and VSSR region are
clearly visible.

A zoom of the optical SDSS 2nd Data Release\footnote{Sloan Didital Sky Survey:
\texttt{http://www.sdss.org/dr2/}} and Chandra maps of the very central region
is displayed in Fig.~\ref{fig:optX}. An X-ray point source which may be an AGN
is visible southeast of the cluster center. The distribution of the X-ray gas
towards the west and northwest edges of the central region seems somewhat more
clearcut than in the rest of the cluster centre. This suggests a cold front
comparable to that previously found in other clusters, e.g.,
\citet{Markevitch00}. A ``hole'' in X-ray emission is also visible south of
the cluster center.

\subsection{Temperature and metallicity maps}

\begin{figure}[!htb]
\includegraphics[width=\columnwidth]{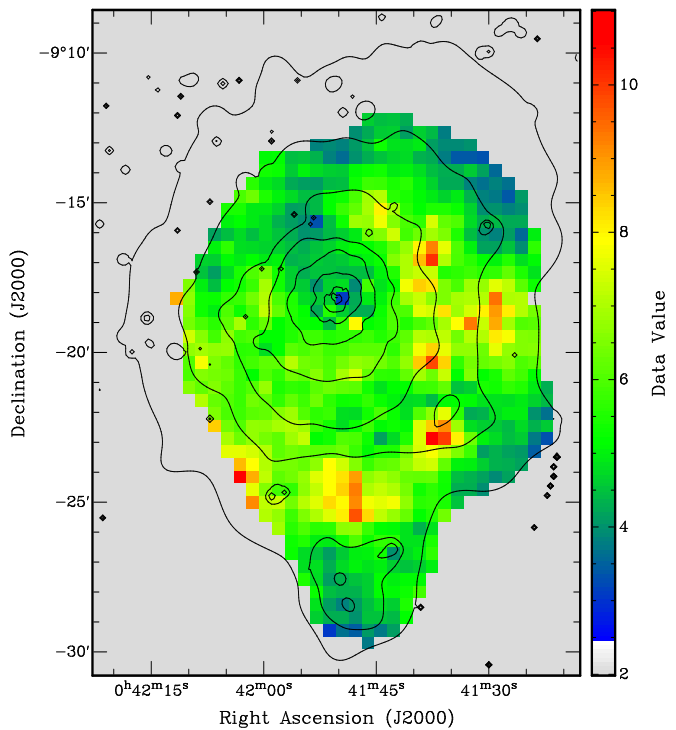}
\caption[]{Temperature map and XMM isocontours in the [0.5-8.0]
keV band for Abell 85.  The scale for the temperature is keV.}
\label{fig:tx}
\end{figure}

\begin{figure}[!htb]
\includegraphics[width=\columnwidth]{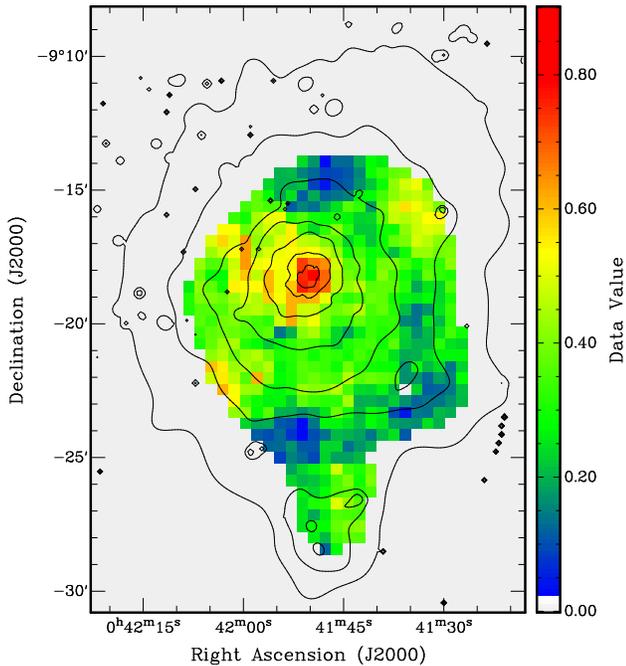}
\caption[]{Metallicity map with XMM X-ray intensity contours in
the [0.5-8.0] keV band superimposed. The scale for the metallicity is
in solar units. }
\label{fig:Zx}
\end{figure}

The temperature and metallicity maps are made in a grid, where each pixel is
$512 \times 512$ XMM EPIC physical pixels, i.e., each cell grid is $25.6''
\times 25.6''$. So, from now on ``1 pixel'' is actually this fat $25.6''
\times 25.6''$ ``pixel''. In each pixel we try a spectral fit to determine
simultaneously the temperature and metallicity.

We set a minimum count number ($\sim 600$ counts for a temperature map,
$\sim 4000$ for a metallicity map) necessary for proceeding with a
spectral fit using the \textsc{mekal} plasma model.

If we do not have the minimum count number in a pixel, we try a square
region of $3 \times 3$ pixels; if we still don't have the minimum
number of counts, we try a $5 \times 5$ pixel region. For the
metallicity map only, we went up to a $7 \times 7$ pixel region. If we
don't have enough counts, the pixel is ignored and we proceed to the
next neighbouring pixel. This is done for all pixels in the grid,
covering most of the MOS image.

When we do have enough counts, the spectral fit is done with the
hydrogen column density fixed at the local galactic value -- for each
pixel, we estimate $N_{\rm H}$ using the task \texttt{nh} from
\textsc{ftools} (which is an interpolation from the \citet{Dickey90}
galactic $N_{\rm H}$ table). Note that we compute the RMF and ARF
matrix for each pixel in the grid.

Since the cluster covers more or less the entire XMM-Newton field, the
background was taken into account by extracting spectra (for MOS1,
MOS2 and PN) from the EPIC blank sky templates described by Lumb et
al. (2002) and publicly available. We have applied the same filtering
procedure to the background event files and extracted the spectra in
all the regions. Finally, the spectra have been rebinned with the
\texttt{grppha} task, to reach at least 20 counts per energy bin. 
We checked that there was no difference in the soft Galactic emission
(in the [0.3-1.0] keV range) between the source and background
files. We also checked that the 10-12 keV count rates were the same in
our data and in the background files.

The temperature map (Fig.~\ref{fig:tx}) reveals that the gas is colder
near the center, as expected from its cooling flow
classification. Several hotter patches are observed on the west half
of the cluster. The fact that the region located just north of the south
blob is hotter agrees with the scenario in which this is the
``impact'' region where the groups forming the filament hit the
cluster, and therefore compress and heat the intracluster gas,
probably through a shock.

The very steep spectrum radio source (VSSR)
\citep{Bagchi98,LimaNeto01} located southwest of the center
corresponds to a somewhat cooler region (see its full spectral
analysis below).

Typical errors on temperatures in this map are about 10\%.

The metallicity map (Fig.~\ref{fig:Zx}) was drawn by setting the
ratios of the heavy element abundances relative to that of iron equal
to the solar ratios.  Regions in grey are those where the spectral
signal to noise ratio was not sufficient to derive a metallicity.  The
central region, with an extension towards the northeast, has a notably
higher metallicity, as expected from the higher density of galaxies
which tend to increase the heavy element abundances in the
intracluster medium through supernova winds. A higher metallicity is
also observed in two zones more or less symmetrically located on
either side of the central region towards the northwest and southeast,
at the extremities of the image. The northwest region has a cooler
temperature, while the southeast region has a hotter temperature, so
the higher metallicities in these two zones cannot be explained by an
anti-correlation between the metallicity and temperature. Typical
errors on metallicities in this map are about 10--15\%.

\section{Radial profiles}

We have determined the radial temperature, metallicity and absorbing
hydrogen column density profiles using the combined MOS1, MOS2 and PN
events. We have used circular annuli, with a width determined by
requiring a minimum of about 2000 counts per annulus; this condition
was obtained by trial and error, requiring a compromise between
spatial resolution and accurate spectral fitting. All three
parameters (temperature, metallicity and absorption) were left free to
vary.

\subsection{Temperature profile}

\begin{figure}[!htb]
\includegraphics[width=8.5cm]{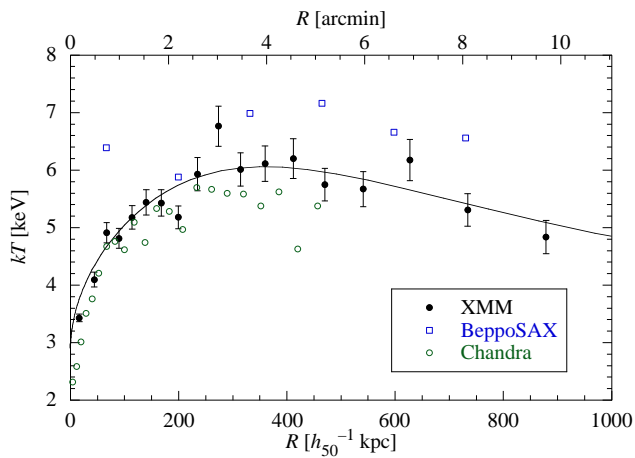}
\caption[]{Gas temperature as a function of radius obtained with
XMM-Newton (filled circles), Chandra (empty circles) and BeppoSax
(empty squares).}
\label{fig:kT_r}
\end{figure}

The XMM-Newton data points for the gas temperature as a function of radius are
shown together with the BeppoSax and Chandra points in Fig.~\ref{fig:kT_r}.
There is an obvious decrease of the gas temperature towards the center for
radii below about 2.5~arcmin (250~kpc). The temperature is constant between
radii of 250 and 450~kpc, then slowly decreases outwards. \textbf{Note that
the overall shape of the temperature profile remains the same (within error
bars) whether we exclude the south blob and VSSR regions or not.}

In order to take the radial variations of the temperature into account
when deriving the dynamical mass, we searched for a mathematical
approximation to the temperature profile.

As a mathematical approximation to the temperature
profile, we fit the following empirical law:
\begin{equation}
    T(R) = T_0 + 2 T_0 \frac{\left(R/r_{t}\right)^{1/2}}{1 + 
    \left(R/r_{t}\right)^{2}} \, ,
    \label{eq:T(r)}
\end{equation}
where $r_{t}$ is a scale parameter and $T_{0}$ is the central temperature.

The best least-squares fit to the data is obtained for $k T_0=2.83 \pm
0.04\,$keV and $r_{t} = (630 \pm 50) h_{50}^{-1}\,$kpc. Fig.~\ref{fig:kT_r}
shows the best fit temperature profile superposed on the data points; the
agreement is fairly good.

We use the profile given by Eq.~(\ref{eq:T(r)}) as an approximation to
the deprojected temperature profile, $T(r)$. For physically realistic
gas models, the difference between the actual (deprojected) gas
temperature and the measured emission weighted temperature should be
smaller than $\sim 10\%$, well within the observational uncertainties
\citep[e.g.,][]{Markevitch99,Komatsu01}. Using Eq.~(\ref{eq:T(r)}) as
an approximation to the true temperature profile allows us to compute
analytically the dynamical mass profile and the corresponding total
(dark matter plus baryon) density profile
(cf. Sect.~\ref{sec:dynMass}).

\subsection{Metallicity profile}

\begin{figure}[!htb]
\includegraphics[width=8.5cm]{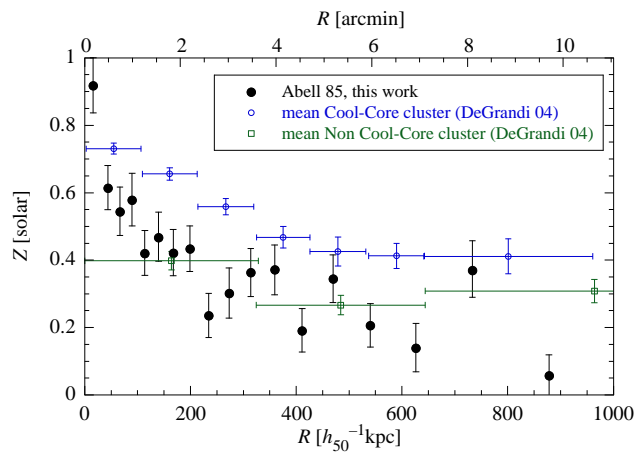}
\caption[]{Gas metallicity profile from our XMM-Newton data. For comparison we 
also plot the \citet{deGrandi04} data for ``cool-core'' and non ``cool-core'' 
clusters.}
\label{fig:ProfileZ}
\end{figure}

The radial metallicity profile is displayed in
Fig.~\ref{fig:ProfileZ}. The metallicity is close to solar in the
innermost radial bin, and decreases to about 0.3Z$_\odot$ for radii
larger than about 350~kpc.

The metallicity profile of Abell~85 has a shape similar to the mean
metallicity profiles obtained by \citet{deGrandi04} for their sample of
``cool-core'' clusters, but with systematically lower values. The metal
abundance of Abell~85 is closer to the non ``cool-core'' values.

\subsection{Absorption profile}

\begin{figure}[!htb]
\includegraphics[width=8.5cm]{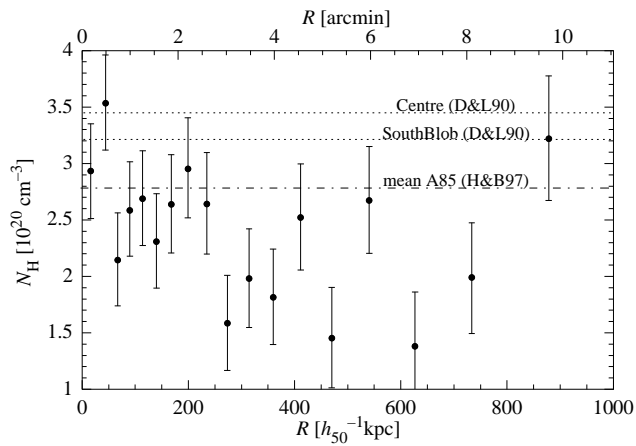}
\caption[]{Hydrogen absorption column profile from our XMM-Newton data. The
two horizontal dotted lines indicate the Galactic absorption columns towards
the Centre of Abell 85 and towards the South Blob obtained with the
\textsc{ftools} \texttt{nh} task, based on \citet[see text for
details]{Dickey90}; the dot-dashed line is the galactic $N_{\rm H}$ value
from \citet{Hartmann97}, cf. \citet{Arabadjis99}.}
\label{fig:ProfileNH}
\end{figure}

The hydrogen absorption column profile is shown in Fig.~\ref{fig:ProfileNH}.
For comparison we also indicate the galactic value derived from
\citet{Dickey90} (using the task \texttt{nh} from \textsc{ftools}) and
\citet{Hartmann97} [using the value quoted by \citet{Hartmann97}]. The values
from Dickey \& Lockman are well above ours, while that of Hartmann \& Burton
is in better agreement with ours.

Though error bars are too large for this effect to be clearly significant, we
can note that the absorption column tends to be smaller than the Galactic
value at almost all radii, as previously found in a totally independent
analysis based on ROSAT PSPC data \citep{Pislar97}. Since the galactic $N_{\rm 
H}$ maps are made on a coarser resolution than the scale at which clusters are 
observed in X-rays, the actual $N_{\rm H}$ in the direction of Abell~85 may be 
smaller than the value quoted in these maps.

The absorption profile is apparently not flat: there may be stronger
absorption in the central region, within 3~arcmin. Such a behaviour
was already noticed with BeppoSAX data by \citet{LimaNeto01} in
Abell~85, and in other clusters \citep[e.g.,][from XMM
data]{Pointecouteau04}.  The origin of such an absoption gradient, if
present, is still controversial \citep[and references
therein]{Allen00,Pointecouteau04}; a possible hypothesis is absorption
due to very cool molecular gas and/or dust grains. Notice however that
the error bars are still too large to confirm the absorption gradient.

\section{Physical properties and metal abundances in various regions}

We now analyze in detail the spectral properties of various regions of
the cluster.  A list of the X-ray point sources excluded in the
various fits is given in Table~\ref{tbl:pointsources}; all the
excluded zones around point sources have radii of 12 arcsec, except
for source 3 which is a particularly bright Seyfert (see below).  We
derive the temperatures and iron abundances, and wherever possible the
abundances of several other heavy elements, without fixing their
ratios to the iron abundance equal to the solar ratios. An obvious
requirement for this is to have enough counts in the spectra. As an
example, there are 63000, 68000, 38000, 12300, 14800 and 5600 PN
counts in the Centre, Ring~1, Ring~2, South blob, VSSR and Impact
regions respectively (after excluding point sources).

\subsection{The Central zone and two concentric rings}

\begin{table}[!htb]
\centering
\caption[]{Point sources excluded in the various spectral fits, sorted by R.A.}
\begin{tabular}{c c c }
\hline
Source number & R.A. (J 2000) & Dec. (J2000) \\
\hline

1  & 00 41 26.354 & -09 22 12.09 \\
2  & 00 41 29.705 & -09 22 45.70 \\
~3$^*$  & 00 41 30.167 & -09 15 47.85 \\
4  & 00 41 36.516 & -09 23 40.12 \\
5  & 00 41 37.273 & -09 22 34.52 \\
6  & 00 41 39.759 & -09 23 17.72 \\
7  & 00 41 40.625 & -09 21 20.92 \\
8  & 00 41 40.841 & -09 22 36.12 \\
9  & 00 41 42.810 & -09 26 25.90 \\
10 & 00 41 43.242 & -09 27 35.07 \\
11 & 00 41 44.539 & -09 28 26.27 \\
12 & 00 41 46.486 & -09 29 33.47 \\
13 & 00 41 48.216 & -09 28 42.27 \\
14 & 00 41 50.150 & -09 25 47.59 \\
15 & 00 41 50.920 & -09 28 19.87 \\
16 & 00 41 53.165 & -09 25 05.71 \\
17 & 00 41 57.490 & -09 24 40.10 \\ 
\hline
$^*$ Seyfert galaxy & & \\
\end{tabular}
\label{tbl:pointsources}
\end{table}

We first analyze three regions drawn in Fig.~\ref{fig:Xray-all},
starting with a circular region of radius $R=100$~arcsec encompassing
the cluster center (hereafter region ``Center''). This corresponds
more or less to the cooling flow zone, where the temperature drops,
and therefore a constant temperature may not be a very good
approximation.

We then extracted two regions corresponding to concentric annuli,
where the temperature is expected to be roughly constant: one between
radii of 100 and 250~arcsec (Ring~1), the other between radii of 250
and 400~arcsec (Ring~2). Two small circular zones were excluded from
Ring~2: one of radius 24 arcsec centered on the Seyfert galaxy (source
3 in Table~\ref{tbl:pointsources}) and one of radius 80 arcsec centered
on the VSSR.  It was not possible to extract zones further out with a
sufficient number of counts to estimate independently the various
heavy element abundances.

For these three regions, we have simultaneously fit the spectra obtained with
all three detectors (MOS1, MOS2 and PN) with the \textsc{vapec} and
\textsc{vmekal} models \citep{Kaastra93,Liedahl95}, using the photoelectric
absorption given by \citet{Balucinska92}. For the central region, we also fit
the spectra with a \textsc{cevmkl} model, which takes into account the existence of a
cooling flow. The energy ranges taken into account were 0.3--10.0 keV for the
two MOS spectra and 0.5--8.0 keV for the PN spectrum. The Galactic absorption
and redshift were left free to vary. For several elements, abundances could
not be estimated, and were frozen to 0.3 times the solar value.

\begin{figure}[!htb]
\includegraphics[width=8.5cm]{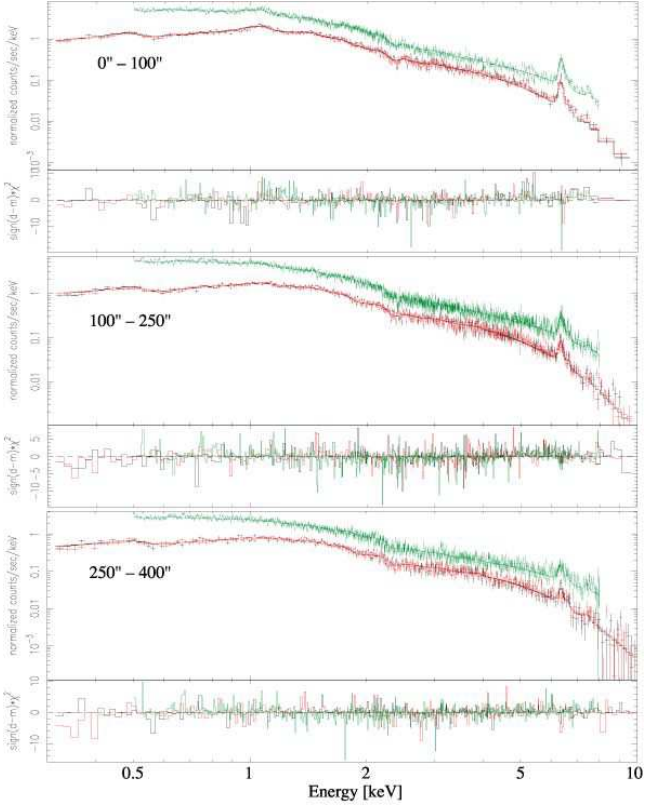}
\caption[]{Best spectral fits with residues for XMM-Newton MOS1
(black), MOS2 (red) and PN (green). Top: central region ($R < 100''$)
fit with a cooling-flow model (\textsc{cevmkl}); middle: Ring~1 fit
with a \textsc{vmekal} ($100'' < R < 250''$); bottom: Ring~2 fit with
a \textsc{vmekal} ($250'' < R < 400''$). }
\label{fig:spectre1}
\end{figure}

Fig.~\ref{fig:spectre1} shows the spectra for regions Center, Ring~1 and
Ring~2, together with the best \textsc{vmekal} or \textsc{cevmkl}
(cooling-flow \textsc{mekal} model) fits and residuals in terms of individual 
$\chi^{2}$ contribution of each energy bin.

Table~\ref{tbl:specfitscen} summarizes the results obtained for the
various fits on these three regions. The iron abundance is estimated
with a typical precision of 3\% in the Center and better than 10\% in
the two Rings. The other elements measured in the central zone are: O,
Ne, Mg, Si, S, Ca and Ni. Mg and S are not detected in the Rings.

It is interesting to note that the abundances of the various metals
relative to solar are not at all the same from one element to another.
In the central region, Fe and O are about 0.3 and 0.4Z$_\odot$ 
respectively, while Ne and Si are somewhat higher and Ni is much higher.

Clear metallicity gradients are also present, all elements but Ca being
notably less abundant in the Rings than in the Center, suggesting
that there is little or no radial mixing or convection of the gas. 

Fluxes and luminosities were calculated based on the best fit
for the three detectors independently. We give in
Table~\ref{tbl:fluxlum} the averages of these quantities in the
[0.5-2.0 keV] and [2.0-10.0 keV] bands obtained from MOS1 and MOS2.
The bolometric X-ray luminosities (integrated in the [0.001-100 keV]
interval) are also given. Typical errors are of the order of 10\%.

\begin{figure}[!htb]
\includegraphics[width=8.5cm]{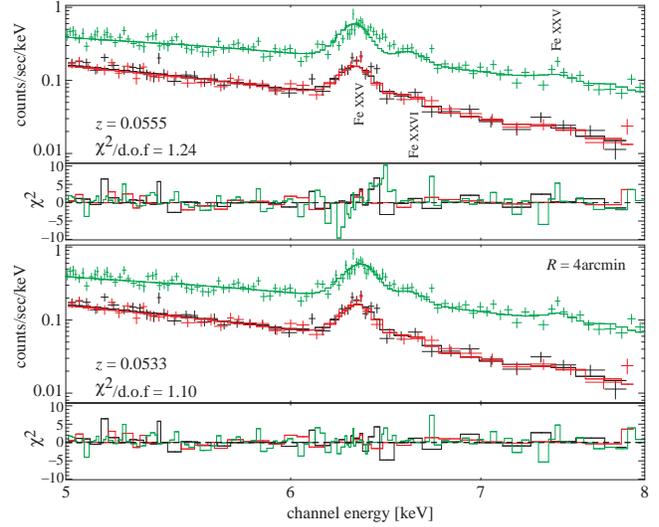}
\caption[]{Spectral \textsc{mekal} fit within the central 4~arcmin using the
three XMM cameras, restricted to the [5.0--8.0 keV] band. Each line
corresponds to one of the 3 detectors, cf. Fig.~\ref{fig:spectre1}. Top panel:
temperature and metallicity are free and the redshift is set to 0.0555. Bottom
panel, when the redshift is set free, the best fit is $z =
0.0533_{-0.0009}^{+0.0015}$ (90\% confidence level). The reduced $\chi^{2}$ is
given in each panel.}
\label{fig:A85_Xredshift}
\end{figure}

Interestingly, the redshift derived from the various X-ray gas fits in the
central region is between 0.0530 and 0.0535, corresponding to velocities of
$cz = 15900$--16050~km/s. Fig.~\ref{fig:A85_Xredshift} illustrates the redshift
dependency on the quality of the fit. It seems clear that the X-ray emitting
gas in the central 4~arcmin is best fit with $z = 0.0533$.

On the other hand, the mean optical redshift of Abell~85 is 0.0555,
corresponding to a velocity of $cz = 16650\,$km/s
\citep{Durret98a}. The mean optical velocity of the cluster is
therefore larger than that of the X-ray gas in the innermost 100
arcsec radius region by $\sim 600$~km/s. This difference is
significant only at a 2$\sigma$ level since the MOS (PN) energy
calibration is accurate to 5 eV (10 eV), translating to a velocity of
250 km/s (500~km/s).  Since Abell~85 is known to show velocity
substructures, we extracted from the velocity catalogue of
\citet{Durret98a} the galaxies in the innermost 100 arcsec radius
region and with velocities in the cluster interval, as defined by
\citet{Durret98b}. Four galaxies correspond to these criteria, and
their mean velocity is $cz=$15903~km/s, a value consistent with the
X-ray gas velocity given above.

Interestingly, the cD appears to have a higher velocity
$cz=$16734~km/s [$z = 0.0558 \pm 0.0002$, \citet{Durret98a}] implying
that it is certainly not resting at the bottom of the cluster
potential well traced by the X-ray gas. This velocity difference is in
the upper range of cD peculiar velocities measured by
\citet{Oegerle01}, in agreement with the general formation scenario
for Abell~85 involving several mergers \citep[e.g.,][]{Dubinski98}.

\begin{table*}[!htb]
\centering
\caption[]{Spectral fits of the central region (radius 100~arcsec)
with three different models, and of regions Ring~1 (radius 100 to
250~arcsec) and Ring~2 (radius 250 to 400~arcsec) with two different
models. Abundances are in solar units and error bars are 1$\sigma$.
Values of $N_H$  are in units of $10^{20}cm^{-2}$.}
\begin{tabular}{l c c c c c c c }
\hline
Region	  & Center                     & Center                      & Center                       & Ring 1                      &Ring 1                       &Ring 2                   & Ring 2  \\
Model     & vmekal                     & cevmkl                      & vapec                        & vmekal                      &vapec                        &vmekal                   & vapec    \\
\hline
$N_H$     &2.50$_{-0.12}^{+0.10}$&2.70$_{-0.02}^{+0.02}$     &2.83$_{-0.002}^{+0.001}$      &1.78$_{-0.15}^{+0.18}$ &1.985$_{-0.001}^{+0.001}$    &1.194$_{-0.002}^{+0.002}$& 1.448$_{-0.002}^{+0.001}$ \\
kT (keV)  & 4.30$_{-0.03}^{+0.04}$     & 5.77$_{-0.15}^{+0.15}$      & 4.18$_{-0.03}^{+0.03}$       & 5.70$_{-0.04}^{+0.09}$      &5.69$_{-0.08}^{+0.05}$       &6.28$_{-0.13}^{+0.10}$   & 6.20$_{-0.14}^{+0.08}$      \\
Redshift  &0.0530$^{-0.0002}_{+0.0002}$& 0.0533$_{-0.0004}^{+0.0001}$& 0.0535$_{-0.0004}^{+0.0004}$ & 0.0527$_{-0.0009}^{+0.0011}$&0.0535$_{-0.0006}^{+0.0010}$&0.0505$_{-0.0009}^{+0.0047}$&0.0539$_{-0.0005}^{+0.0004}$\\
$\chi^{2}$& 1741                       & 1670                        & 1755                         &  1770                        & 1766                       & 1437                    & 1427                        \\
dof       & 1491                       & 1490                        & 1491                         &  1636                       &1635                         & 1346                    & 1332                        \\
\hline
O  & 0.25$^{-0.05}_{+0.06}$    & 0.30$_{-0.05}^{+0.05}$    & 0.38$_{-0.06}^{+0.09}$    & 0.022$_{-0.022}^{+0.089}$  &0.20$_{-0.12}^{+0.11}$    & 0.32$_{-0.15}^{+0.18}$  & 0.25$_{-0.19}^{+0.24}$ \\
Ne & 0.53$^{-0.09}_{+0.10}$    &         \nodata           & 0.78$_{-0.10}^{+0.15}$    & 0.32$_{-0.11}^{+0.12}$     &0.44$_{-0.12}^{+0.19}$    &0.75$_{-0.19}^{+0.18}$   & 1.01$_{-0.23}^{+0.27}$ \\
Mg &          \nodata          & 0.18$_{-0.10}^{+0.10}$    & 0.17$_{-0.12}^{+0.13}$    &     \nodata                &    \nodata               &0.29$_{-0.24}^{+0.24}$   &    \nodata             \\
Si & 0.55$^{-0.06}_{+0.07}$    & 0.60$_{-0.06}^{+0.06}$    & 0.56$_{-0.07}^{+0.08}$    & 0.16$_{-0.09}^{+0.09}$     &0.15$_{-0.10}^{+0.10}$    &       \nodata           &    \nodata             \\
S  & 0.27$^{-0.07}_{+0.09}$    & 0.35$_{-0.08}^{+0.08}$    & 0.28$_{-0.09}^{+0.09}$    &     \nodata                &     \nodata              &       \nodata           &    \nodata             \\
Ca & 0.35$^{-0.21}_{+0.24}$    & 0.37$_{-0.24}^{+0.26}$    & 0.46$_{-0.23}^{+0.21}$    & 0.56$_{-0.31}^{+0.31}$     &0.57$_{-0.33}^{+0.29}$    &0.79$_{-0.50}^{+0.56}$   & 1.09$_{-0.58}^{+0.55}$ \\
Fe & 0.408$^{-0.013}_{+0.018}$ & 0.409$_{-0.011}^{+0.012}$ & 0.451$_{-0.014}^{+0.014}$ & 0.271$_{-0.013}^{+0.012}$  &0.296$_{-0.015}^{+0.012}$ &0.236$_{-0.021}^{+0.021}$& 0.261$_{-0.020}^{+0.024}$ \\
Ni & 1.39$^{-0.25}_{+0.24}$    & 1.54$_{-0.24}^{+0.24}$    & 1.25$_{-0.29}^{+0.19}$    & 0.46$_{-0.28}^{+0.27}$     &     \nodata              &1.34$_{-0.49}^{+0.54}$   & 1.37$_{-0.45}^{+0.75}$ \\
\hline
\end{tabular}
\label{tbl:specfitscen}
\end{table*}

\begin{table*}[!htb]
\centering
\caption[]{Spectral fits of the South Blob (circle of radius 200~arcsec),
VSSR (circle of radius 150~arcsec) and Impact regions, with two different
models. Abundances are in solar units and error bars are 1$\sigma$.
Values of $N_H$  are in units of $10^{20}cm^{-2}$.}
\begin{tabular}{l c c c c c c}
\hline
Region	  & South Blob & South Blob & VSSR  & VSSR & Impact & Impact \\           
Model     & vmekal      & vapec     & mekal & apec & mekal  & apec   \\
\hline
$N_H$     & 0.042$_{-0.042}^{+0.046}$ & 0.042$_{-0.041}^{+0.044}$ & 0.93$_{-0.29}^{+0.30}$ & 1.00$_{-0.30}^{+0.30}$ & 0.147$_{-0.147}^{+0.548}$ & 0.082$_{-0.082}^{+0.088}$ \\  
kT (keV)  & 5.74$_{-0.26}^{+0.26}$ & 5.79$_{-0.27}^{+0.25}$ & 6.26$_{-0.21}^{+0.22}$ & 6.23$_{-0.20}^{+0.20}$ & 8.87$_{-0.65}^{+0.67}$ & 8.84$_{-0.60}^{+0.69}$ \\ 
Redshift  &0.0555 & 0.0555 & 0.0555 & 0.0555 & 0.0555 & 0.0555  \\ 
$\chi^{2}$& 775 & 774 & 916 & 917 & 504 & 502 \\ 
dof       & 758 & 758 & 882 & 882 & 533 & 533 \\
\hline
O  & 0.26$_{-0.19}^{+0.24}$ & 0.40$_{-0.30}^{+0.36}$ &\nodata &\nodata &\nodata &\nodata \\ 
Ne & 0.75$_{-0.30}^{+0.35}$ & 1.11$_{-0.43}^{+0.49}$ &\nodata &\nodata &\nodata &\nodata \\ 
Mg & 0.52$_{-0.40}^{+0.46}$ & 0.68$_{-0.52}^{+0.52}$ &\nodata &\nodata &\nodata &\nodata \\
Si & 0.25$_{-0.23}^{+0.28}$ & 0.35$_{-0.29}^{+0.31}$ &\nodata &\nodata &\nodata &\nodata \\ 
Fe & 0.168$_{-0.031}^{+0.033}$ & 0.190$_{-0.036}^{+0.037}$ & 0.201$_{-0.034}^{+0.035}$ & 0.223$_{-0.038}^{+0.039}$ & 0.220$_{-0.084}^{+0.085}$ & 0.246$_{-0.094}^{+0.095}$ \\ 
\hline
\end{tabular}
\label{tbl:specfitsreg}
\end{table*}

\begin{table*}[!htb]
\centering
\caption[]{X-ray fluxes and luminosities in regions Center, Ring~1 and Ring~2.
Fluxes are in units of $10^{-11}$ erg cm$^{-2}$ s$^{-1}$ and luminosities in units of
$10^{44}$ erg s$^{-1}$.}
\begin{tabular}{l c c c c c c}
\hline
Region	             & Center  & Ring 1 & Ring 2 & South Blob & VSSR & Impact \\  
\hline
Flux [0.5-2.0 keV]   & 1.35    & 1.29   & 0.65   & 0.19       & 0.22 & 0.070\\
Flux [2.0-10.0 keV]  & 1.81    & 2.12   & 1.14   & 0.29       & 0.38 & 0.145\\
\hline						           	
L$_X$ [0.5-2.0 keV]  & 1.88    & 1.81   & 0.91   & 0.25       & 0.31 & 0.10\\
L$_X$ [2.0-10.0 keV] & 2.67    & 3.10   & 1.65   & 0.42       & 0.54 & 0.21\\
L$_{X,bol}$          & 6.17    & 6.64   & 3.51   & 0.92       & 1.17 & 0.44\\
\hline
\end{tabular}
\label{tbl:fluxlum}
\end{table*}

\subsection{The South blob}

\begin{figure}[!htb]
\includegraphics[width=8.5cm]{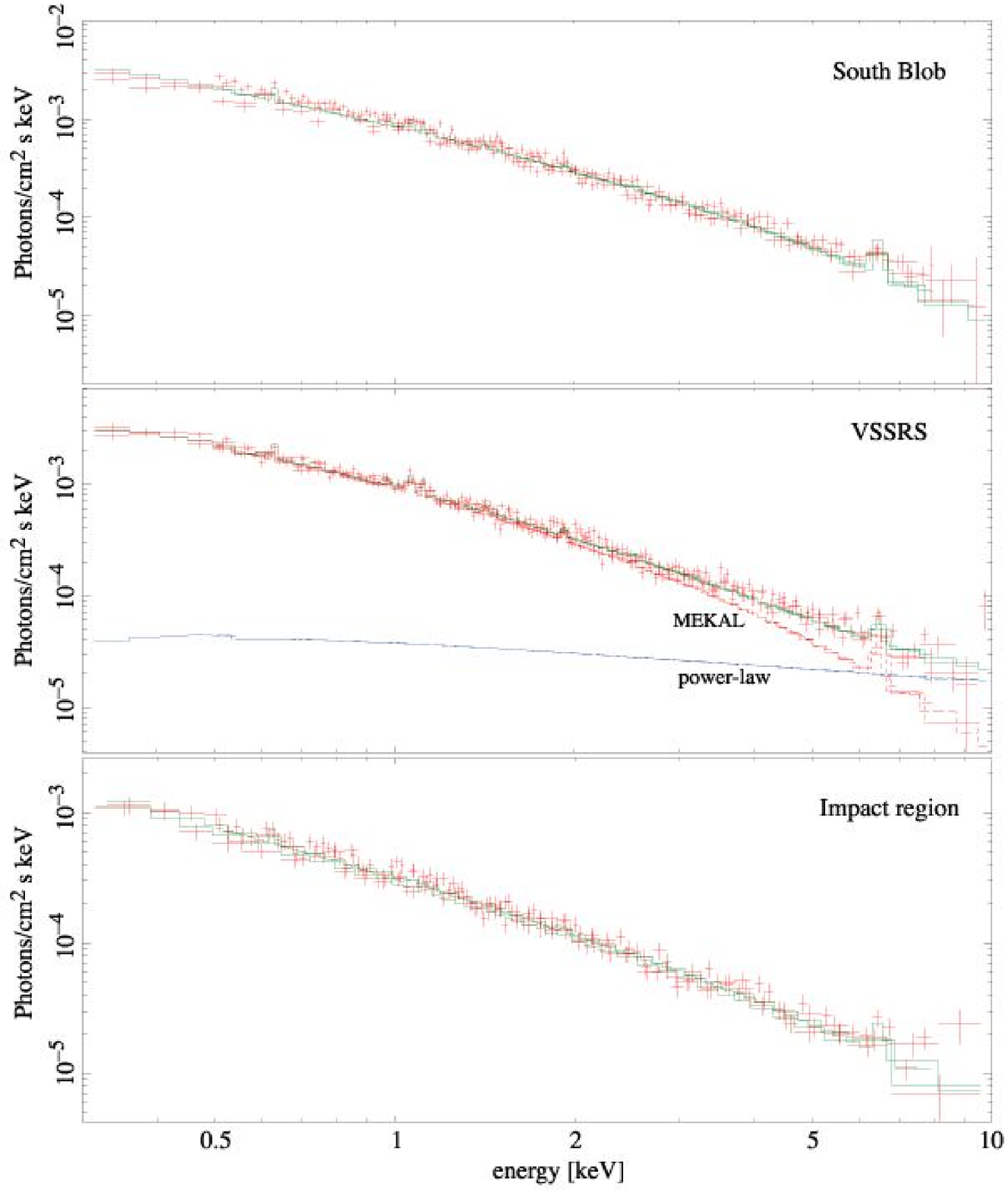}
\caption[]{Photon spectra for 3 individual regions. Top: the VSSRS region
(2.5~arcmin radius region). Middle: the elliptical impact region. Bottom: the 
South Blob region (3.0~arcmin radius region). See text for region locations.}
\label{fig:specblobS}
\end{figure}

The South blob spectrum was obtained using the Chandra image to derive
the positions of the point sources, which are not all clearly visible
on the XMM-Newton image, and exclude them from the XMM-Newton spectrum
extracted for the South blob in a 3.0~arcmin radius region. It was
then fit by \textsc{mekal} and apec spectra. The corresponding
spectrum and \textsc{vmekal} fit are shown in the top panel of
Fig.~\ref{fig:specblobS}, and results are given in
Table~\ref{tbl:specfitsreg}. The temperature is kT=$5.75\pm 0.25$ keV.
Contrary to the regions previously considered, where the redshift is
left free to vary in the fit, the result here is not acceptable, in
view of the mean optical redshift, which is found to correspond to a
velocity of 16826 km/s (based on 16 galaxy redshifts). We therefore
fixed the redshift to the mean cluster value of 0.0555.

\subsection{The VSSR region}

\begin{figure}[!htb]
\includegraphics[width=8.5cm]{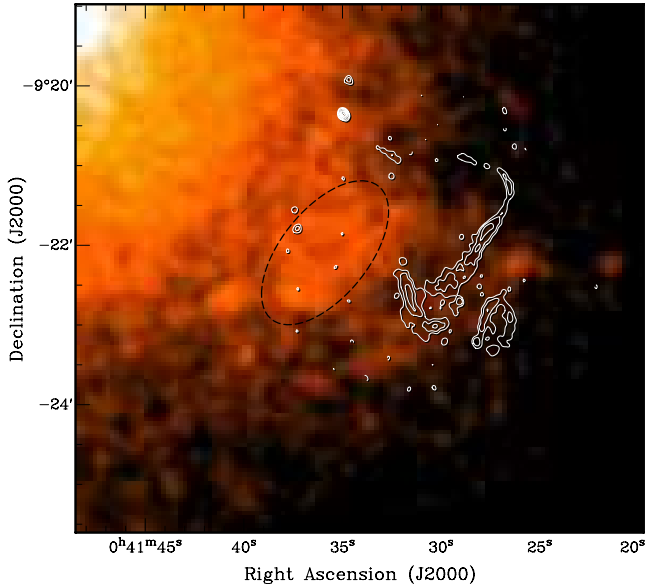}
\caption[]{MOS1+MOS2 X-ray image with the 21cm radio isocontours
obtained by \citet{Slee01} overlayed. The radio source seems to be
anti-correlated with the west X-ray excess emission at $\alpha \sim
0^{\rm h}41^{\rm m}36^{\rm s}$ and $\delta \sim -9^\circ 22'$
approximately limited by the dashed ellipse.}
\label{fig:SleeRadio}
\end{figure}

The region called VSSR corresponds to the very steep spectrum radio source
MRC~0038-096 \citep{Large91}. Fig.~\ref{fig:SleeRadio} shows the high
resolution 21cm VLA radio isocontours of \citet{Slee01} superposed on the
MOS1+MOS2 X-ray image.

This radio source was thought to be associated with the south-west
X-ray excess already observed from ROSAT data
\citep{LimaNeto97,Bagchi98}.  \citet{Slee01}, using higher resolution
radio data, also associated the radio source to the X-ray excess based
on ROSAT data. With XMM data, however, the relation between the X-ray
excess and the radio data is not very clear, but we can see from
Fig.~\ref{fig:SleeRadio} that the high resolution VLA radio source
does not coincide spatially with the X-ray excess. Such a phenomenon
could happen if the relativistic electrons blow away some of the X-ray
emitting gas \citep{Boehringer93,Fabian00}.

A spectrum was extracted in a circular region of radius 2.5~arcmin
corresponding to the VSSR (see Fig.~\ref{fig:Xray-all}). The best
thermal fit is obtained for $kT=6.2\pm 0.20 $~keV. We also tried a double
component fit, including a thermal component and a power law. The
power-law component is intended to represent the inverse Compton
emission of cosmic microwave background photons on the relativistic
electrons responsible for the synchrotron radio emission observed in
this location \citep{Bagchi98}. However, the non-thermal component is
not unambiguously disentangled from the thermal emission, and the
reduced $\chi^{2}$ is not significantly better. Assuming that we have
detected a non-thermal component (with the above caveat), this
non-thermal X-ray flux is $1.5 \times
10^{-12}~$erg~s$^{-1}$cm$^{-2}$. The middle panel of
Fig.~\ref{fig:specblobS} shows the best fit spectrum.

There are 8 galaxies in the VSSR region with measured redshifts, but
the corresponding velocities spread between 13781 and 17800~km/s with
no concentration at a specific velocity, so we cannot extract any
kinematical information.

\subsection{The impact region}

As mentioned above, the region believed to be undergoing the impact of the
infalling groups on to the main cluster is located between the south blob and
the cluster center. We extracted the MOS1+MOS2+PN spectrum in an ellipse of
major axis 2.43~arcmin, minor axis 1.33~arcmin and major axis position angle
aligned with the east-west direction (see Fig.~\ref{fig:Xray-all}). 

The spectrum of this region is shown in the bottom panel of
Fig.~\ref{fig:specblobS} and the corresponding results for a
\textsc{mekal} and an apec model are given in
Table~\ref{tbl:specfitsreg}. As in the two previous cases, if the
redshift was left free to vary in the X-ray spectral fit, we found
unreasonably small values, so we had to fix the redshift to 0.0555. As
expected from the temperature map, the temperature in this zone is
notably higher than in the rest of the cluster: $kT=8.8\pm 0.6$~keV,
in agreement with the idea that the gas in that region is compressed
and heated by a shock due to the infall of groups on to the main
cluster.

\subsection{Metallicity gradients}

\begin{figure}[!htb]
\includegraphics[width=8.5cm]{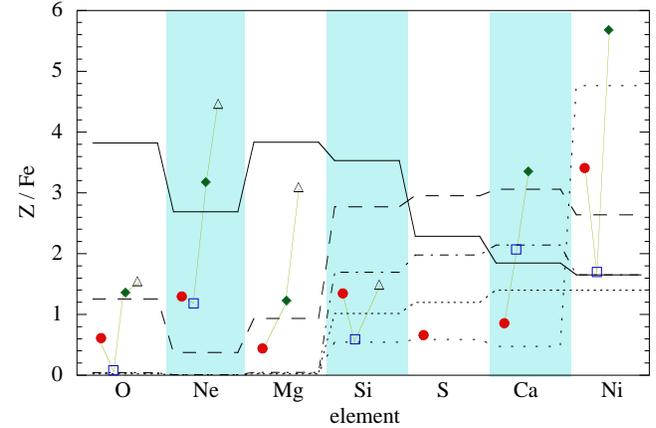}
\caption[]{Heavy element abundances relative to iron in regions Centre
(red filled circles), Ring~1 (empty squares), Ring~2 (green filled
diamonds) and South Blob (empty triangles). Various models are
superimposed: SNII (full line), Hypernovae (long dashes), and SNIa
WDD1 (dot-dashed), WDD2 (dotted) and W7 (spaced dots).}
\label{fig:metals}
\end{figure}

The ratios of the heavy element abundances relative to iron in regions Center,
Ring~1, Ring~2 and South blob are shown in Fig.~\ref{fig:metals}. Various
theoretical metal yield models are superimposed: ``Hypernovae''
\citep{Nakamura01}, SN~II, and SN~Ia models WDD1, WDD2 and W7
\citep{Nomoto97}.

In none of the four regions does any single model account for the
heavy element abundances. However, Fig.~\ref{fig:metals} shows that in
the Centre and Ring~1 SNIa seem capable of reproducing the observed
abundances. This may be explained by the presence of a central cD
galaxy which, as normal ellipticals, is deficient in type II SNe but
may have a relatively high SNIa rate -- roughly 16 type Ia SN per
century \citep{Cappellaro01}.

On the other hand, the high Ne, Ca and Ni abundances measured in Ring~2
require the contribution of SNII -- this could have happened during the
initial burst of star formation within the clusters elliptical galaxies
\citep[e.g.,][]{Matteucci95,Pepino02}. Note that Ring~2 is dominated by the
region towards the northwest of the cluster which has a notably stronger
metallicity. Ne and Mg in the South Blob are also closer to those due to SNII
models while Si is consistent with SNIa models. Therefore, it does not seem
possible from our data to estimate accurately the proportion of SNIa and SNII
responsible for the enrichment of the intracluster medium, as in
\citep[e.g.][]{Dupke00a,Dupke00b,Sanders04}; besides being a function of time,
this proportion is probably a function of position (or local density) in
the cluster \citep{Tamura04}.

\section{Mass determination}

\subsection{Gas mass}

The gas density distribution is determined from the radial X-ray
surface brightness profile, $\Sigma_{X}$, using the fact that:
\begin{equation}
    \Sigma_{X}(R) \aproxsim \int_{R}^{\infty} n_{e}^{2}(r) \frac{r \mbox{d} 
    r}{\sqrt{r^{2} - R^{2}}} \, ,
    \label{eq:rho-sigma}
\end{equation}
where $n_{e}$ is the electron number density of the intracluster gas. For an 
electrically neutral gas, $n_{e} \approx n_{\rm H}$ and then $\rho(r) \approx 
m_{p} n_{e}(r)$, where $m_{p}$ is the proton mass.

We have used two analytical profiles to describe $\Sigma(R)$, the S\'ersic 
(\citeyear{Sersic68}) law:
\begin{equation}
    \Sigma(R) = \Sigma_{0} \exp\left[- \left(\frac{R}{a}\right)^{\nu}\right] 
    \, ,
    \label{eq:Sersic}
\end{equation}
and the $\beta$-model \citep{Cavaliere76}:
\begin{equation}
    \Sigma(R) = \Sigma_{0} \left[1 + 
    \left(\frac{R}{R_{c}}\right)^{2}\right]^{-3\beta +1/2}
    \, ,
    \label{eq:betamodel}
\end{equation}
where $a$ and $R_{c}$ are the scale parameters, $\nu$ and $\beta$ the shape 
parameters, and $\Sigma_{0}$ the intensity normalization.

\begin{figure}[!htb]
\includegraphics[width=8.5cm]{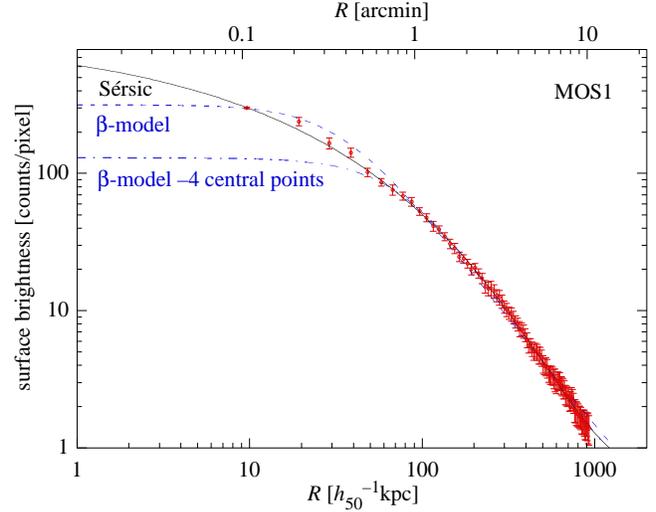}
\caption[]{Emission profile of the X-ray gas obtained from the
XMM-Newton MOS1 image (the profile from MOS2 is litterally indistinguishable 
from MOS1); three fits are superimposed: a S\'ersic law
(black), a $\beta$-model (blue dashes) and a $\beta$-model excluding
the four innermost points (blue dot-dashes).}
\label{fig:sx}
\end{figure}

The X-ray brightness profile of Abell 85 was obtained with the
STSDAS/IRAF task \textsc{ellipse}, using the XMM-Newton MOS1 exposure
corrected [0.3--8.0~keV] band image. We have masked the CCD gaps and
point sources. In Fig.~\ref{fig:sx}, we show the brightness profile
together with three different fits: a S\'ersic profile, a
$\beta$-model and a $\beta$-model excluding the four innermost points,
which correspond to a known excess emission with respect to a flat core
\citep[cf.][]{Gerbal92,LimaNeto97}.

If all the points are included, the S\'ersic law provides a better fit than
the $\beta$-model; however, if the four innermost points are excluded, the
S\'ersic and $\beta$-model become indistinguishable.

We have chosen the S\'ersic profile to describe the emissivity profile
for its simplicity and capability to describe well the whole profile 
\citep[see e.g.][]{Demarco03}. There is an analytical approximation
to the deprojection of a S\'ersic profile \citep{LimaNeto99}. Thus, if
the brightness profile is given by Eq.~(\ref{eq:Sersic}), the electron
number density is:
\begin{equation}
    \begin{array}{l}
        n_{e}(r) = n_{0} (r/a)^{-p} \exp\left[- (r/a)^{\nu} \right]\, ;\\[3pt]
         2 p = 1 - 0.6097 \nu + 0.05563 \nu^{2} \, ,
    \end{array}
    \label{eq:MMmodif}
\end{equation}
and, for Abell 85, a least-squares fit of the brightness profile gives $a
= 35 \pm 7\,$kpc, $\nu = 0.39 \pm 0.02$ and $p = 0.39 \pm 0.03$.

In order to estimate the electronic density, $n_{0}$ [the intensity
normalization in Eq.~(\ref{eq:MMmodif})], we integrate the bremsstrahlung
emissivity along the line-of-sight within 4~arcmin in the central region and
compare the result with the flux obtained by spectral fitting of the same
region (the normalization parameter of the thermal spectral model in
\textsc{xspec}, which is proportional to $n_{e}^{2}$). We thus obtain $n_{0} =
(77.0 \pm 6.5) \times 10^{-3}\,$cm$^{-3}$.

\begin{figure}[!htb]
\includegraphics[width=8.5cm]{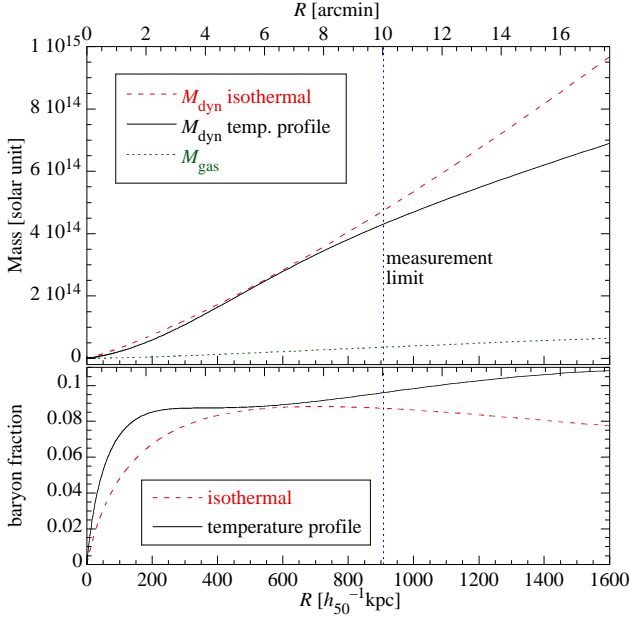}
\caption[]{Top: gas mass (green dotted line) and dynamical mass (black
solid line when including the calculation of the temperature profile,
red dashed line in the isothermal case) as a function of radius.  The
vertical blue dotted line indicates the limiting radius of the data.
For the isothermal case, we have adopted the mean emission temperature $kT = 
6.5\,$keV.
Bottom: baryon fraction in both cases.}
\label{fig:mdynmgas}
\end{figure}

The gas mass as a function of radius can then simply be derived by
integrating the gas density in spherical shells:
\begin{equation}
    M(<r) = \frac{4 \pi a^{3} n_{0} m_{p}}{\nu}\  
    \gamma\left[\frac{3-p}{\nu},\left(\frac{r}{a}\right)^{\nu}\right] \, ,
    \label{eq:massGas}
\end{equation}
where $\gamma(z,x) \equiv \int_{0}^{x} e^{z} z^{t-1} {\rm d} t$ is the
incomplete gamma function. We use $\rho(r) = m_{p} n_{e}(r)$, supposing the
gas is electronically neutral and $m_{p}$ is the proton mass. The gas mass is
shown in Fig.~\ref{fig:mdynmgas}.

At the limiting radius of our data, $r=10$~arcmin or about $900 h_{50}$~kpc,
the gas mass is $3.5 \times 10^{13} M_{\odot}$.

\subsection{Dynamical mass and baryon fraction}\label{sec:dynMass}

In order to compute the dynamical mass, the temperature profile is introduced.
Assuming hydrostatic equilibrium and spherical symmetry, the S\'ersic profile 
described above and the empirical temperature profile, Eq.~(\ref{eq:T(r)}), 
the dynamical mass is:
\begin{equation}
    \begin{array}{l}
     M_{\rm dyn}(<r) =  kT_{0} r \, \times 3.7 \times 10^{10} M_{\odot} 
     \times \\[3pt]
     \displaystyle{\frac{
    \left[p + \nu \left(x \frac{r_{t}}{a} \right)^{\nu} \right]
    (1 + x^{2}) \left(1 + 2 \sqrt{x} + x^{2}\right) -
    \sqrt{x} + 3 x^{5/2}}{(1 + x^{2})^{2}}} \, ,
    \end{array}
    \label{eq:Mdyn}
\end{equation}
where $x \equiv r/r_{t}$, $a$ and $r_{t}$ are in kpc and $kT_{0}$ in keV.

If we assume an isothermal temperature profile, the dynamical mass reduces 
to:
\begin{equation}
    M_{\rm dyn}^{\rm iso} = 3.7 \times 10^{10} kT r \left[p + \nu 
    \left(\frac{r}{a}\right)^{\nu}\right]\, M_{\odot},
    \label{eq:MdynIso}
\end{equation}
with $r$ and $a$ in kpc.

The dynamical mass is shown in Fig.~\ref{fig:mdynmgas} as a function of
radius, both for isothermal and non-isothermal gas temperatures. For
the isothermal case, we have adopted the mean emission weighted temperature
$kT = 6.5\,$keV. These two curves are very similar within the limiting radius
of our data, but start to diverge further out, showing that even when the
temperature does not vary very strongly with radius (see Fig.~\ref{fig:kT_r}),
it has a non negligible influence on the total mass estimate at large radius.
At the last measured point, $r = 10\,$arcmin, $M_{\rm dyn} = 4.3 \times
10^{14} M_{\odot}$ and $M_{\rm dyn}^{\rm iso} = 4.7 \times 10^{14} M_{\odot}$.

The dynamical and gas masses allow us to estimate the baryon fraction
profile in Abell~85. The mass in baryons is computed assuming that
$\sim 16$\% is contributed by the galaxies, i.e., $M_{\rm baryon}
\approx 1.16 M_{\rm gas}$ \citep{White93}.

The bottom panel in Fig.~\ref{fig:mdynmgas} shows the baryon fraction
for both isothermal and non-isothermal temperature profiles. It is
seen to increase strongly with radius within the inner 150~kpc, then
to remain roughly constant. This agrees with previous findings that dark
matter is strongly peaked towards the centers of clusters
\citep[e.g.][]{Gerbal92,Allen01}, in agreement with numerical
simulations (e.g. Rasia et al. 2004). The baryon fraction at $r =
10\,$arcmin is 0.09 and 0.10 for isothermal and non-isothermal
profiles, respectively.

\subsection{Virial radius and halo density}

Having the dynamical mass we can compute the radius $r_{\delta}$ that 
corresponds to 
\begin{equation}
    \overline{\rho(r_{\delta})}/\rho_{c}(z) = \delta \, ; \quad
    \overline{\rho(r_{\delta})} \equiv \frac{M_{\rm dyn}}{4\pi 
    r_{\delta}^{3}/3} \, ,
    \label{eq:rdelta}
\end{equation}
where $\rho_{c}(z)$ is the critical density of the Universe at redshift $z$. 
For the SCDM and $\Lambda$CDM models, the virial radius corresponds to
$\delta = 180$ and 340 respectively  \citep[cf.][]{Lacey93,Bryan98}. 

Eq.~(\ref{eq:rdelta}) cannot be solved analytically when we take into
account the temperature profile and the dynamical mass given by
Eq.~(\ref{eq:Mdyn}), but it can be solved numerically and thus we have:
$$
\begin{array}{rccccc}
  \delta :                             &  180 &  200 & 340 & 500 & 2500 \\
  r_{\delta} [h_{50}^{-1}\,\mbox{kpc}] : & 2697 &  2563& 1995& 1671& 789
\end{array}
$$
Errors on $r_{\delta}$ are about 15\% ($1\sigma$ confidence level). Taking the
virial radius as $r_{\rm vir} = 2 h_{50}^{-1}\,$Mpc, the radial range where we
are able to obtain a temperature profile for Abell~85 is $0.005 \la r/r_{\rm
vir} \la 0.45$.

\begin{figure}[!htb]
\includegraphics[width=8.5cm]{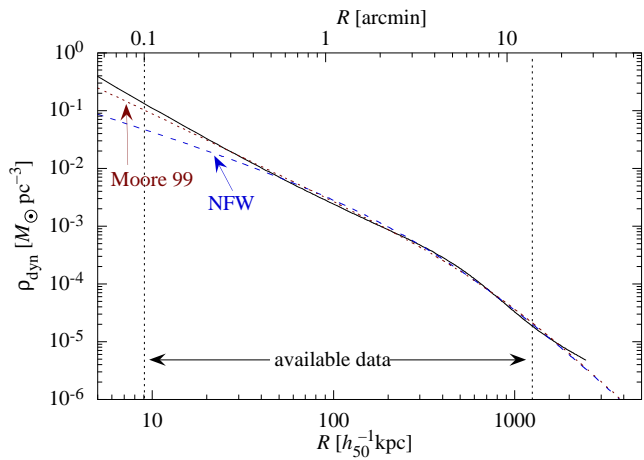}
\caption[]{Dynamical mass density as a function of radius derived from
our data (full line) and compared to a NFW model (dashed line) and to
a Moore99 profile (dotted line).  $\rho_{dyn}\propto r^{-1.9}$ is a
good fit at small radii. The two vertical dotted lines indicate the
spatial resolution (left) and the limit of our data (right).}
\label{fig:rhodyn}
\end{figure}

We also derived the dynamical mass density as a function of radius, that can
be compared to the density of dark matter halos formed in cosmological N-body
simulations. In Fig.~\ref{fig:rhodyn} we show the dynamical mass density 
compared to the \citet[``NFW'']{NFW97} ``universal'' profile for 
dark halos and to the \citet[``Moore99'']{Moore99} profile.

In the interval $0.1 < r < 10\,$arcmin, where the temperature and
surface brightness are reliably determined, both theoretical profiles
agree quite well (the Moore99 profile slightly better) with the
density profile we obtained for Abell~85.

The characteristic radius of the NFW profile that best fits our
density profile is $r_{\rm NFW} = 390 h_{50}^{-1}$ kpc; with the above
determined $r_{\rm vir}$, this implies that the concentration
parameter is $c = r_{\rm vir}/r_{\rm NFW} \approx 5$. This value is in
agreement with the cosmological N-body simulations of \citet{Jing00},
in a $\Lambda$CDM scenario.

\section{Discussion and conclusions}
\label{discussion}

Our XMM-Newton data have allowed us to obtain temperature and
metallicity maps for Abell~85 and to derive heavy element abundances
in various regions.

The temperatures obtained for the Center and Rings (see Table
\ref{tbl:specfitscen}) agree very well with those measured by
\citet{Markevitch98} from ASCA data in his regions~1 and 6--9
respectively. On the other hand, \citet{Donnelly03} find a fairly
homogeneous temperature throughout Abell~85, except in a small zone
southeast of the center, and in a large one towards the northwest
edge. This difference may be due to the lower spatial resolution and
sensitivity of ASCA.

The temperatures and/or metallicities of several regions seem to
indicate that Abell~85 has undergone or is still undergoing merging
events. However, there is no intensity enhancement whatsoever in any
of the hotter or more metal rich regions, suggesting that there is no
gas density discontinuity.

We can note that the SE region is the only place where $kT$ and $Z$
are not anti-correlated: it is hot and metal rich; however this zone
is just at the limit where metallicities can be measured, so we
will not take it into account in the coming discussion.

Numerical simulations of a cluster merger such as those displayed,
e.g., by \citet{Bourdin04} in their Fig.~9, show that when two
clusters merge the gas temperature is not affected immediately, but
the final temperature map can look strikingly similar to the one
we have derived, with various hotter zones.  Different numerical
simulations have also shown that star formation activity in galaxies
was only briefly enhanced by a merger, maybe less than 3~Gyr,
according to Fig.~4 in \citet{Fujita99}, but then due to ram pressure
stripping galaxies saw their star formation rate (SFR) decrease. These
two sets of simulations can help us propose a possible scenario
that could account for what we are seeing in Abell~85.

First, there appears to be a ``filament'' extending at least 4~Mpc
southeast of the cluster and probably made of groups falling on to the
main cluster \citep{Durret03} and hitting it in the Impact region.  In
favour of this interpretation are two main facts: i)~the temperature
of the filament is about 2.0 keV, consistent with that of groups;
ii)~the impact region, located more or less half way between the
cluster center and the south blob, is notably hotter (8.8 keV). This
is most probably the place where a particularly large group hits the
main cluster.

Second, the NW region is cooler and more metal rich than the average
cluster. A cooler temperature is expected in metal-rich regions, but
the fact that this zone is more metal rich probably implies that it
has undergone star formation with a higher rate than the rest of the
cluster, possibly triggered by a group merging from the NW towards the
SE. Note however that the NW zone, as the NE one, is just at the limit
where metallicities can be measured, so we should take these results
with caution.

Third, the NE region has no temperature enhancement, but has a notably
higher metallicity, here also suggesting metal enrichment due to
enhanced star formation and/or ram pressure stripping due to a merger.

Fourth, there are hot zones distributed as an arc from south to
NW. These could also be the result of a major merger coming from the
NW, since as mentioned above a strong similarity is observed between
our X-ray temperature map and that of the numerical simulations of a
cluster merger by \citet{Bourdin04}.

If a cluster has fallen on to Abell~85 from the NW, this would be
consistent with the existence of a metal rich zone in the NW, due to a
brief increase of the SFR which produced a metal enrichment of the
ICM, followed by a decrease in the SFR, in agreement with the
homogeneity of the metallicity map in the rest of the cluster. Such a
scenario suggests that this merger has taken place more than 4~Gyr ago
in order for the SFR to be presently reduced \citep{Fujita99}. If we
follow the same line to interpret the NE zone, where the gas has a
high metallicity but a temperature similar to the average cluster
temperature, this would require a second merger coming from the NE,
but notably more recent (less than 3~Gyr), so that star formation is
still triggered, thus enhancing the ISM metallicity, but no effects
are visible yet on the temperature. In this scenario, the impact
region due to the infall of the ``filament'' mentioned above would
then be a relatively old merger.

Such a scenario appears to be consistent with the various heavy
element abundances measured in the Center and in regions Ring~1 and
Ring~2, which imply chemical enrichment of the intergalactic gas by
both supernovae of type Ia and of type II, and corresponding to
different star formation rates.

Therefore, the interpretation of the temperature and metallicity maps
derived for the X-ray gas in Abell~85 based on numerical simulations
by \citet{Fujita99} and \citet{Bourdin04} leads us to suggest that
Abell~85 has undergone three mergers, two rather old ones (older than
4~Gyr) from the NW and South, and a more recent one from the NE.  Of
course, this is only a schematic scenario since it neglects galaxy
movements within the cluster, and assumes that galaxies enrich the ISM
at their own location, through enhanced star formation and/or ram
pressure stripping. As pointed out by the referee, we should also be
aware that repeated AGN outbursts dredging up material from the center
could also produce comparable effects.  More detailed modeling
obviously requires a better knowledge of the galaxy distribution, in
particular of emission line galaxies.  We hope to achieve this shortly
through multi band imaging allowing to derive spectroscopic redshifts
and H$\alpha$ imaging to draw the distribution of galaxies with high
star formation rates.

\begin{acknowledgements}

We thank the referee, Dr J.C. Kempner, for several interesting
suggestions that helped improve our analysis.  F.D. and
G.B.L.N. acknowledge financial support from the USP/COFECUB and from
the PICS France-Br\'esil from CNRS. G.B.L.N. acknowledges support from
FAPESP, CNPq and EARA.
This work is based on observations obtained with XMM-Newton, an ESA
science mission with instruments and contributions directly funded by
ESA Member States and the USA (NASA).

\end{acknowledgements}

\end{document}